\documentclass[pre,aps,amsmath,amssymb,amsfonts,floatfix,superscriptaddress,showpacs,twocolumn,footinbib]{revtex4-1}

\usepackage[colorlinks=true,citecolor=blue,urlcolor=blue,linkcolor=blue,hyperfigures=true]{hyperref}
\usepackage{graphicx}
\usepackage{amsmath,amsfonts}
\usepackage[dvipsnames]{xcolor}
\usepackage{color}
\usepackage{soul}

\usepackage{inputenc}
\usepackage{comment}

\newcommand{\up}{\uparrow}
\newcommand{\dn}{\downarrow}

\newcommand{\kv}{\ensuremath{\mathbf{k}}}

\newcommand{\qv}{\ensuremath{\mathbf{q}}}
\newcommand{\Qv}{\ensuremath{\mathbf{Q}}}

\newcommand{\sz}{\ensuremath{\text{sp}}}
\newcommand{\glat}{\Gamma}
\newcommand{\gimp}{\Gamma_\text{loc}}

\renewcommand{\sl}{} 
\newcommand{\FK}{}

\usepackage[normalem]{ulem}

\newcommand{\last}[1]{{\color{black} #1}}

\def\presuper#1#2%
  {\mathop{}%
   \mathopen{\vphantom{#2}}^{#1}%
   \kern-0.5\scriptspace%
   #2}

\usepackage{tikz}
\usetikzlibrary{calc}
\usetikzlibrary{arrows}
\usetikzlibrary{arrows.meta}
\usetikzlibrary{patterns}
\usetikzlibrary{decorations.pathmorphing}
\usetikzlibrary{decorations.pathreplacing}
\usetikzlibrary{decorations.markings}
\usetikzlibrary{shapes.geometric}
\usetikzlibrary{positioning}

    \tikzset{middlearrow/.style={
                decoration={markings,
                            mark= at position 0.65 with {\arrow{#1}},
                                    },
                                            postaction={decorate}
                                                }
                                                }
    \tikzset{middlearrowpart/.style={
                decoration={markings,
                            mark= at position 0.65 with {\arrow{Triangle}},
                                    },
                                            postaction={decorate}
                                                }
                                                }
                                                
    \tikzset{middlearrowhole/.style={
                decoration={markings,
                            mark= at position 0.65 with {\arrow{Triangle[open,fill=white]}},
                                    },
                                            postaction={decorate}
                                                }
                                                }

\begin{document}

\author{Friedrich Krien}
\email{krien@ifp.tuwien.ac.at}
\affiliation{Institute for Solid State Physics, TU Wien, 1040 Vienna, Austria}

\author{Paul Worm}
\affiliation{Institute for Solid State Physics, TU Wien, 1040 Vienna, Austria}

\author{Patrick Chalupa-Gantner}
\affiliation{Institute for Solid State Physics, TU Wien, 1040 Vienna, Austria}

\author{Alessandro Toschi}
\affiliation{Institute for Solid State Physics, TU Wien, 1040 Vienna, Austria}

\author{Karsten Held}
\affiliation{Institute for Solid State Physics, TU Wien, 1040 Vienna, Austria}

\title{Explaining the {pseudogap} {through damping and antidamping\\ on the Fermi surface by imaginary spin scattering}}

\begin{abstract}
\noindent {\textbf{Abstract}}
    The mechanism of the pseudogap observed in hole-doped cuprates remains one of the central puzzles in condensed matter physics.
    We analyze this phenomenon via a Feynman-diagrammatic inspection of the Hubbard model. Our approach captures the pivotal interplay between Mott localization and Fermi surface topology {\sl beyond} weak-coupling spin fluctuations, which {would} open a spectral gap near hot spots. We show that strong coupling and particle-hole asymmetry {trigger a very different} mechanism: {a large imaginary part of the spin-fermion vertex promotes damping of antinodal fermions }{and, at the same time, protects the nodal Fermi arcs (antidamping). Our analysis naturally explains puzzling features of the {pseudogap} observed in experiments, such as Fermi arcs being cut off at the antiferromagnetic zone boundary and the subordinate role of hot spots.}
\end{abstract}

\maketitle

\noindent {\textbf{Introduction}}\\
The single-band Hubbard model is believed to capture key physics of the cuprates~\cite{Anderson1987,Zhang88} and nickelates~\cite{Li2019,Karp2020,Kitatani2020}. Various numerical and theoretical approaches show that this model exhibits the so-called pseudogap phase~\cite{Kampf90,Pines93,Abanov03,Senechal04,Maier05,Civelli05,Kyung06,Macridin06,Haule07-2,Katanin09,Ferrero09,Kuchinskii12,Gull13,Efetov13,Gunnarsson15,Gunnarsson16,Rohringer16,Chen17,Wu17,Wu18,Scheurer18,Rohringer18,Maier19,Robinson19,Reymbaut19,Qin21}, an extreme nodal/antinodal dichotomy of the Fermi surface (FS), where spectral weight is concentrated on Fermi arcs~\cite{Shen05,Kanigel06,Sobota21}.

{However,} {the precise mechanism responsible for the pseudogap remains one of the most controversially debated topics {in condensed matter physics}.} {On general grounds,} the important role played by spin fluctuations~\cite{Kampf90,Pines93,Abanov03,Vilk97,Vilk97-2,Gunnarsson15} is naturally suggested by the proximity to an antiferromagnetic phase. In the {conventional}, weak-coupling picture a spectral gap opens near hot spots~\cite{Vilk97,Vilk97-2,Wu18}, which is observed in electron-doped cuprates~\cite{Armitage01,Kyung04}. For hole-doped cuprates this is not the case, {instead}, the gap opens near the antinodes~\cite{Kanigel06}, and a reconstruction of the FS~\cite{Eberlein16,Wu18,Scheurer18,Sachdev18} is evidenced by quantum oscillations~\cite{DoironLeyraud07}. {Other features not explained by weak-coupling spin fluctuations are the good Fermi-liquid properties of underdoped cuprates~\cite{Mirzaei13} and indications of broken time-reversal symmetry~\cite{Kaminski02,Xia08,Zhao16}. Alternative origins of the pseudogap~\cite{Anderson04,Yang06,Lee06,Keimer15,Punk15,Imada16} are hence under consideration.}

\begin{figure} [b]
\begin{center}
\begin{tikzpicture}
\node[anchor=south west,inner sep=0,opacity=0.85] (image1) at (0,0) {\includegraphics[scale=0.8]{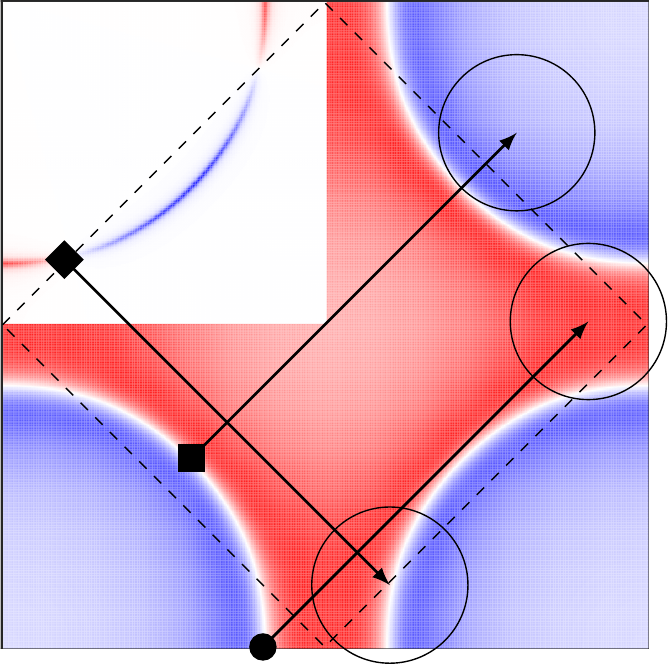}};
\end{tikzpicture}
\end{center}
\vspace{-.5cm}
    \caption{{Top left quadrant: damping (red) and antidamping (blue) on the Fermi surface (FS). Right and bottom quadrants: real part of the noninteracting Green's function. Blue (red) color indicates particle-like (hole-like) states above (below) the Fermi level. Filled symbols: antinode (circle), node (square), hot spot (diamond). Arrows represent the antiferromagnetic wave vector $\Qv$. Open circles $\sim 1/\xi$ comprise available target states; red (blue) states are occupied (unoccupied) and promote damping (antidamping). Dashed lines show the antiferromagnetic zone boundary (AZB)~\cite{umklapp}.}}
\label{fig:bz}
\end{figure}

{Here we unravel the physical origin of the discrepancy between the {conventional} picture of spin fluctuations on the one hand, and both experiments {on hole-doped cuprates} and numerical investigations of the single-band Hubbard model on the other.} {In particular, we unveil} {the strong-coupling \last{spin-fluctuation} mechanism responsible for the pseudogap sketched in the top left quadrant of Fig.~\ref{fig:bz}: spin fluctuations diminish lifetimes of quasiparticles near the antinodes (red), } {while they even {\sl enhance}} {lifetimes near the nodes (blue). The antiferromagnetic zone boundary (AZB, dashed) marks the crossover {between these opposite behaviors}.} {Remarkably, }{the strong-coupling mechanism has no effect near {\sl hot spots}, defined through the intersection of AZB and Fermi surface. Here only the conventional weak-coupling mechanism for spin scattering~\cite{Kampf90,Pines93,Abanov03,Vilk97,Vilk97-2,Wu18} is active. Its effect is however too small to open a gap due to a short antiferromagnetic correlation length} {of} {less than $1$ or $2$ lattice spacings.}

Compared to conventional weak-coupling theory, the effective interaction between spin fluctuations and fermions-----the {\sl spin-fermion vertex}, $\glat$-----plays a radically different role. {At weak coupling $\glat$ is real-valued, which promotes scattering between states `on shell', i.e., close to the Fermi surface. This constraint is ideally fulfilled for hot spots (e.g., filled diamond in Fig.~\ref{fig:bz}), which are connected to other hot spots through the antiferromagnetic wave vector, here $\Qv=(\pm\pi,\pm\pi)$ [arrows]. As spin fluctuations extend over a correlation length $\xi$, the transferred momentum can deviate from $\Qv$ in a circle $\sim 1/\xi$. For large $\xi$ this weak-coupling mechanism opens a gap beginning with the hot spots, {in evident disagreement with experiments on hole-doped cuprates}.}

{In the past, it was reported~\cite{Huang05,Huang06,vanLoon18} that for strong coupling, and if particle-hole symmetry is broken~\cite{Pickem20}, $\glat$ acquires a large imaginary part.} {However, neither Refs.~\cite{Huang05,Huang06} nor, to our knowledge, any previous work noted the {\sl crucial link} between this quantity and the pseudogap at strong coupling.} {Here, based on calculations for the Hubbard model with high spatial resolution, combined with analytic considerations, we identify the imaginary part of $\glat$ as the key to the pseudogap dichotomy.}

{Remarkably, this quantity effectively lifts the nesting condition for spin scattering, allowing fermions to be scattered into off-shell states. Fig.~\ref{fig:bz} shows that antinodal and nodal (filled square) fermions can be scattered into high-energy states far from the Fermi surface. However, the overall feedback on the self-energy depends on the occupancy of the target states: antinodal (nodal) fermions are predominantly scattered into hole-like (particle-like) states, marked with red (blue) color in Fig.~\ref{fig:bz}. As we will show, this increases (diminishes) the scattering rate {at the origin}. We refer to this dichotomy as damping (red) and antidamping (blue). Near hot spots these effects cancel and hence only the weak-coupling mechanism, represented by the real part of $\glat$, is active {in their vicinity}.}\newline

\noindent {\textbf{Method and Model}}\\
    Diagrammatic extensions~\cite{Rohringer18} of dynamical mean-field theory (DMFT)~\cite{Georges96} have proven useful to study spin fluctuations in strongly correlated systems. To reduce bias (see supplementary note 1) we employ the method of Ref.~\cite{Krien20}, corresponding to the parquet approximation~\cite{Bickers04} for dual fermions~\cite{Rubtsov08,Astretsov20}. Through the boson-exchange formalism~\cite{Krien19-4,Bonetti21} we establish a relationship to the spin-fermion model. We apply this machinery to the hole-doped Hubbard model, $H=-\sum_{\langle ij\rangle\sigma}{t}_{ij} c^\dagger_{i\sigma}c^{}_{j\sigma}+ U\sum_{i} n_{i\up} n_{i\dn}.$ Here, $c^\dagger_{i\sigma}$  ($c^{}_{i\sigma}$) create (annihilate) an electron with spin $\sigma$ at site $i$; $n_{\sigma}=c^\dagger_{\sigma}c^{}_{\sigma}$. The nearest $t=1$, next-nearest $t'=-0.2t$, and next-next-nearest $t''=0.1t$ neighbor hopping parameters, and interaction $U=8t$ correspond to  Bi$_2$Sr$_{2-x}$La$_x$CuO$_6$~\cite{Nicoletti10}.\newline

\noindent {\textbf{Results}}\\
\noindent {\textbf{Spin-fermion self-energy.}} To illustrate the mechanism in the spirit of fluctuation diagnostics~\cite{Gunnarsson15,Gunnarsson16}, we consider the following ansatz for the contribution of spin fluctuations with an energy $\omega$ and momentum {\bf q} to the self-energy, $\Sigma_{\sz}(k,q)\propto -G_{k+q}W_q\glat_{kq}.$ Here, $k\!=\!(\kv,\nu), q\!=\!(\qv,\omega)$ are momentum-energy four-vectors, $W_q=-U-\frac{1}{2}U\chi_qU$ denotes the (real-valued) screened interaction, $\chi_q$ the spin susceptibility\last{, $\nu$ ($\omega)$ denote fermionic (bosonic) Matsubara frequencies}. To obtain the full self-energy $\Sigma_{\sz}(k)$ due to spin fluctuations one still has to sum over momenta $\qv$ and frequencies $\omega$. In addition, the full self-energy contains also a momentum-independent contribution $\Sigma^\text{loc}(\nu)$ due to strong local correlations, i.e., $\Sigma(k)=\Sigma^\text{loc}(\nu)+\Sigma_{\sz}(k)$. In the following we consider only the dominant static $q_0=(\qv,\omega=0)$ contribution to the imaginary part of the self-energy,
\begin{align}
\Sigma''_\sz(k,q_0)\propto-&[G''_{k+q_0}\glat'_{kq_0}+G'_{k+q_0}\glat''_{kq_0}]W_{q_0}.\label{eq:spinfermion}
\end{align}
Crucially, $\glat$ has a real ($\glat'$) and an imaginary part ($\glat''$).

    \noindent {\textbf{Imaginary part of the spin-fermion vertex.}} Let us start by considering the conditions for a sizable $\glat''$ in the simpler case of the Anderson impurity model (AIM). We denote its local spin-fermion vertex as $\glat_{\rm loc}(\nu,\omega)$. The leading vertex correction due to local spin exchange has the imaginary part~\cite{Krien19-3} (see supplementary note 2),
\begin{align}
\glat_{\rm loc}''(\nu,\omega\!=\!0)\approx-\frac{TU^2}{2} \chi^\sz_{\omega=0} g'(\nu) g''(\nu),\label{eq:loc_exch}
\end{align}
where $\chi^\sz$ is the spin susceptibility, $T$ the temperature, $g'$ and $g''$ denote real and imaginary part of the impurity Green's function. \FK{Equation~(\ref{eq:loc_exch}) can be considered as a local one-loop correction to the spin-fermion vertex. Let us emphasize that this expression only illustrates how the imaginary part of the vertex arises. The dual fermion numerical results presented later do not rely on this approximation, nor on the restriction to the zeroth bosonic frequency.}

Sufficient for a large $\glat_{\rm loc}''(\imath\eta,\omega\!=\!0)$ \last{[with small positive $\eta$]} are the following conditions: (i) strong particle-hole asymmetry ($g'$ vanishes at symmetry), (ii) large enough spectral weight $-g''(\imath\eta)/\pi$, (iii) large $\chi^\sz_{\omega=0}$ ({\sl preformed} local moment). All of these conditions are satisfied by the DMFT solution of the Hubbard model in the relevant parameter regime for hole-doped cuprates. In general the vertex $\glat$ in the Hubbard model depends on momenta, however, as our numerical calculations below show, the here outlined conditions remain relevant for a large $\glat''$.

\noindent {\textbf{Effect on lifetime.}} We analyze Eq.~\eqref{eq:spinfermion} and put the qualitative considerations regarding Fig.~\ref{fig:bz} on mathematical grounds. {The real part of the vertex, $\Gamma'$, universally {\sl enhances} the magnitude of the imaginary part of the self energy (the scattering rate)~\cite{Pines93,Vilk97,Rohringer16,Wu18,Schaefer21}: the  corresponding term in Eq.~\eqref{eq:spinfermion} is always negative: $-G''\glat'W<0$, since $G''\!<\!0$, $\glat'\!\sim\!1\!>\!0$, and $W\!<\!0$.}

On the contrary, the sign of $-G'\glat''W$ in Eq.~\eqref{eq:spinfermion} depends on the {target} state {with momentum $\kv+\qv$}. We find in our calculations that $\glat''(\kv,\nu,\qv,\omega\!=\!0)$ is an odd function of $\nu$ and in the parameter regime for hole-doped cuprates $\glat''<0$ for Matsubara frequency $\nu>0$. However, the sign of the real part of the Green's function differs: $G'(\kv+\qv,\imath\eta)=[\mu-\varepsilon_{\kv+\qv}-\Sigma'({\kv+\qv},\imath\eta)]/[(\mu-\varepsilon_{\kv+\qv}-\Sigma'({\kv+\qv},\imath\eta))^2+\Sigma''(\kv\last{+\qv},\imath\eta)^2]$  $<0$ ($>0$) for particle-like (hole-like)  target states {shown in blue (red)} color in Fig.~\ref{fig:bz}. We set $\nu>0$, hence
\begin{align}
-G'_{k+q_0}\glat''_{kq_0}W_{q_0}\label{eq:signreg}
\begin{cases}
<0&\text{if $\kv+\qv$ is hole-like,}\\
>0&\text{if $\kv+\qv$ is particle-like.}
\end{cases}
\end{align}
The former enhances the electronic scattering at the Fermi level (damping), the latter {\sl diminishes} it (antidamping). This dichotomy resembles a chemical bonding, where the hybridization with a virtual state at higher (lower) energy reduces (enhances) the energy of the initial state~\cite{Godby88}. The difference is that, due to the complex vertex, this now becomes a dichotomy for the state's lifetime (not its energy).

\begin{figure}
\begin{center}
  \begin{tikzpicture}
\node (image1) at (0,0){\includegraphics[width=0.48\textwidth]{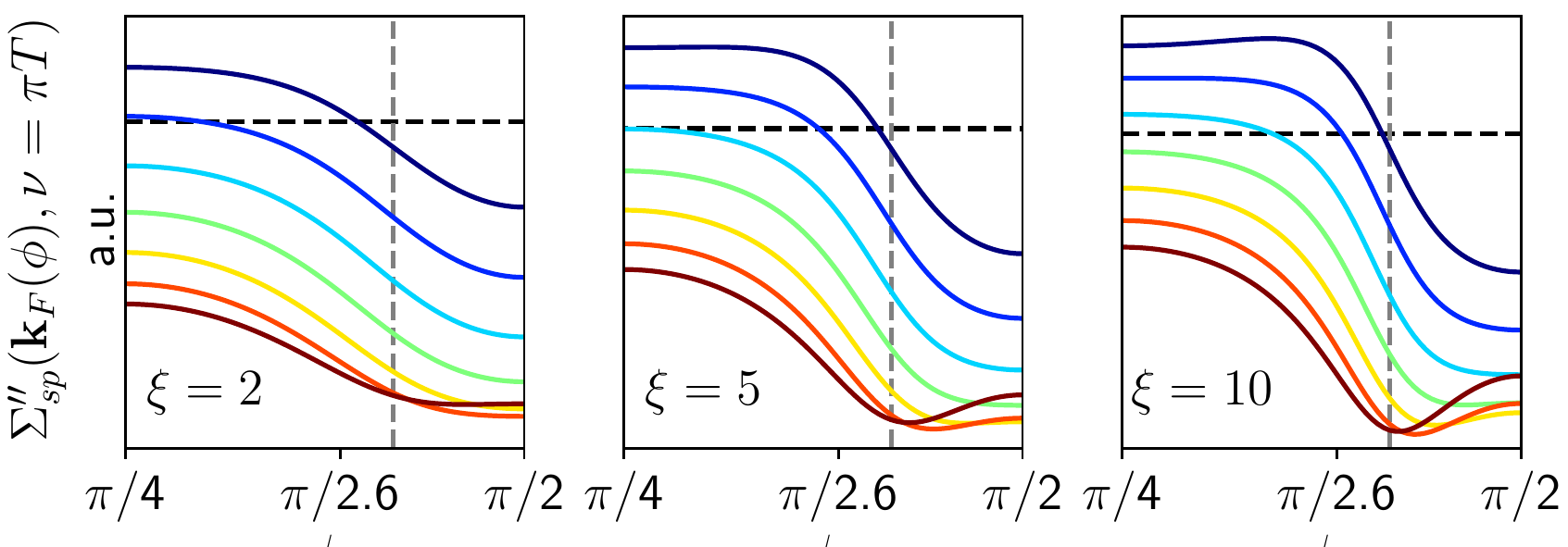}};
\end{tikzpicture}
\end{center}
\vspace{-0.7cm}
    \caption{Model self-energy~\eqref{eq:sfmodel} at the first Matsubara $(\pi T)$ and Fermi vector $\kv_F$, parameterized by the angle $\phi$ from $\pi/4$ (nodal direction) to $\pi/2$ (antinodal direction) for various values of the vertex $\glat_c=e^{\imath\kappa}$. The complex phase is turned from $\kappa=0$ (brown) to $\kappa=-\frac{\pi}{2}$ (dark blue) in steps of $\frac{\pi}{12}$. Panels correspond to the different  correlation lengths $\xi$ as indicated. Vertical lines show the hot spot.}
\label{fig:sfmodel}
\end{figure}

\noindent {\textbf{Relevance to hole-doped cuprates.}} By applying this reasoning to the parameter regime of underdoped cuprates~\cite{Nicoletti10}, {the nodal/antinodal dichotomy of the pseudogap observed in angle-resolved photoemission spectroscopy~\cite{Sobota21} can be explained as follows.} Colors in {the right and bottom quadrants of Fig.~\ref{fig:bz}} indicate the sign of $G'(\kv, \nu\!  = \!\pi/5)$, where we use the noninteracting Green's function $G^0$ at a suitable filling. Let us consider {node, antinode, and hot spot} marked with filled square, circle, and diamond, respectively. Through the vector $\Qv=(\pi,\pi)$ and circles $\sim1/\xi$ we identify the available {target} states. {Red} color of the {target} state (hole, $G'>0$) corresponds to {{\sl damping}}. {Blue} color (particle, $G'<0$) corresponds to {\sl antidamping}. Hence, according to Eq.~\eqref{eq:signreg}, Fermi arcs inside the AZB are cooled, while correlation effects are enhanced on the outside~\cite{umklapp}. Near {hot spots} positive and negative contributions roughly cancel.

\begin{figure}
\begin{center}
\begin{tikzpicture}
\node (image1) at (0,0){\includegraphics[width=0.48\textwidth]{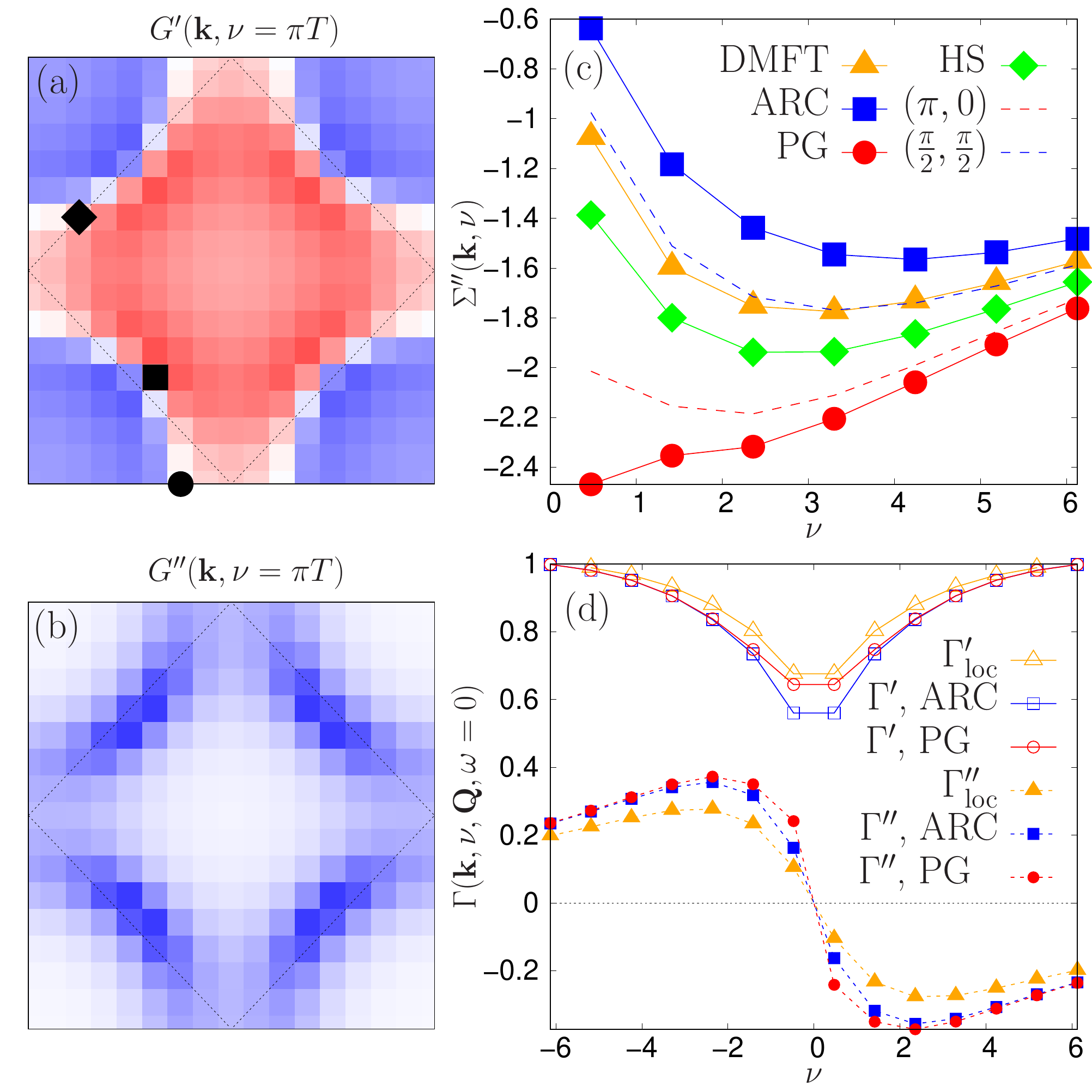}};
\end{tikzpicture}
\end{center}
\vspace{-0.5cm}
    \caption{(a) Real and (b) imaginary part of the Matsubara Green's function $G$ for doping $\delta=0.01$ (\last{here, and in all figures,}~interaction $U=8t$, next- and next-next-nearest neighbor hopping $t'=-0.2t$, $t''=0.1t$, respectively). The correlation length is $\xi\approx1.6$. Black symbols mark Fermi arc (ARC), pseudogap (PG), and hot spot (HS) momentum on the FS. (c) Self-energy and (d) static spin-fermion vertex $\Gamma(\Qv, \omega=0)$ at these and further fermionic momenta as a function of the Matsubara frequency $\nu$. Triangles show local dynamical mean-field (DMFT) quantities.}
\label{fig:pg}
\end{figure}

\noindent {\textbf{Semi-analytical model self-energy.}} The mechanism for the pseudogap due to $\glat''$ is superimposed with the conventional one based on $\glat'$ and the outcome depends qualitatively on the ratio $\glat''/\glat'$. To describe this interplay in a minimal model we define the ansatz \last{(spin-fluctuations in the static, $\omega=0$ limit)} 
\begin{align}
\Sigma_{\sz}(\kv,\last{\imath\eta})\propto \glat_c\frac{T}{N}\sum_\qv\frac{G^0(\kv+\qv,\last{\imath\eta})}{(\Qv-\qv)^2+\xi^{-2}},\label{eq:sfmodel}
\end{align}
where $G^0$ is the noninteracting Green's function shown in Fig.~\ref{fig:bz}. \last{For simplicity, we use here $\eta=\pi T$, which does not affect results qualitatively.} \last{Further, $\glat_c=e^{\imath\kappa}$ is} a complex number with phase $0\geq\kappa\geq-\pi/2$ and of unit length, $T/t=1/5$, and $N$ the number of lattice sites. We restrict the discussion to $\omega=0$ as before, 
and assume the Ornstein-Zernike form for $\chi_q$ peaked around $\Qv=(\pi,\pi)$ with correlation length $\xi$. For simplicity we consider only $\glat_c\equiv\glat_c(\last{\imath\eta})$. {Here, we rotate $\glat_c=e^{\imath\kappa}$ in the complex plane by an angle $\kappa$ away from  $\glat_c=1$ ($\kappa=0$) which corresponds to weak coupling~\cite{Wu18}.}

Fig.~\ref{fig:sfmodel} shows $\Sigma''_{\sz}(\kv_F(\phi),\last{\imath\eta})$ along the FS parameterized by $\phi=\arctan(k_y/k_x)$ from the nodal direction to  the antinodal direction, with increasing correlation  length. Brown lines show the result for $\kappa=0$ $(\glat_c=1)$, which is always negative and for large enough $\xi$ develops a minimum near the hot spot ($\phi_{\text{HS}}\approx1.31$), as expected. Dark blue lines show the result for $\kappa=-\frac{\pi}{2}$ $(\glat_c=-\imath)$ where real and imaginary part of the weak-coupling self-energy are essentially interchanged. Evidently, for suitable $\xi$ and $\kappa$ the minimum of $\Sigma''_{\sz}$ lies at $\phi=\frac{\pi}{2}$, i.e., a gap first opens in the antinodal direction instead of the hot spot. At the same time a finite $\glat_c''=\sin(\kappa)<0$ can lead to positive values of $\Sigma''_{\sz}$ for angles $\phi<\phi_\text{HS}$: clearly, for large $\glat_c''$ the ansatz~\eqref{eq:sfmodel} is meaningful only as a {\sl correction} to a negative $\Sigma^\text{loc}(\nu)$, representing local correlations. 
That is, non-local spin fluctuations {\em enhance} the lifetime of Fermi arcs. 
\last{In supplementary note 2 we compare Eq.~\eqref{eq:sfmodel} to our numerical results.}

\begin{figure}
\begin{center}
\begin{tikzpicture}
\node (image1) at (0,0){\includegraphics[width=0.47\textwidth]{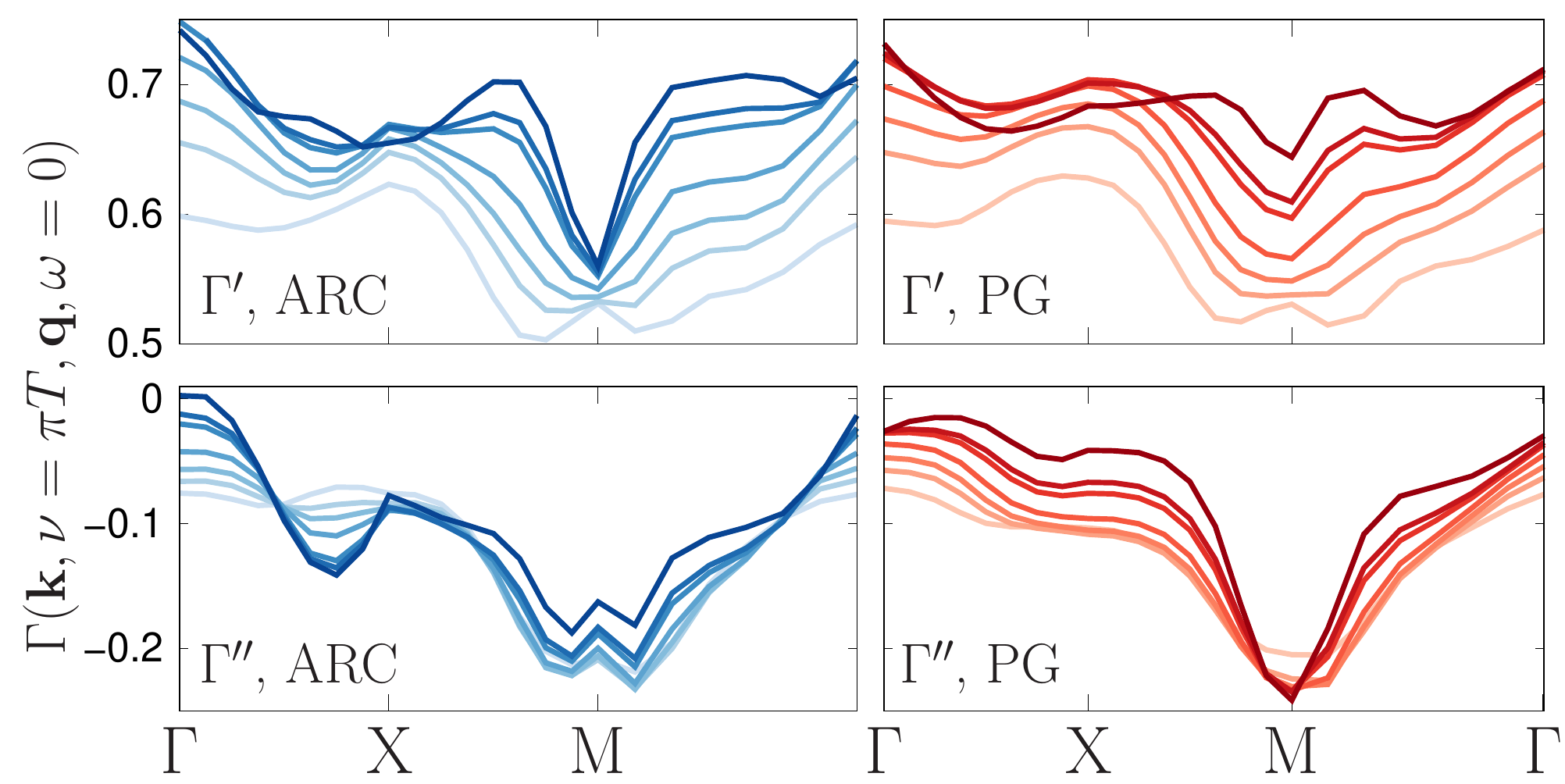}};
\end{tikzpicture}
\end{center}
\vspace{-.5cm}
\caption{Real and imaginary part of the spin-fermion vertex for ARC and PG as a function of $\qv$. Shadings from dark to light: $\delta=0.01, 0.05, 0.07, 0.12, 0.16, 0.2, 0.27$.}
\label{fig:gamma_q}
\end{figure}

\noindent {\textbf{Numerical Results}}\\
We apply the parquet solver for dual fermions presented in Ref.~\cite{Krien20} and evaluate the self-energy $\Sigma_k$ and the (dual) spin-fermion vertex $\glat_{kq}$. The dual formulation implies some more specific features addressed in the supplementary note 1, but the physical interpretation of $\glat$ is consistent with the discussion above. We fix the lattice size to $N=16\times16$, the temperature is $T=0.15t$. {Below we refer to node, antinode, and hot spot as `ARC', `PG', and `HS', respectively.}

Fig.~\ref{fig:pg} (a,b) show the Green's function in the pseudogap phase at doping $\delta=0.01$. The structure of $G'$ is consistent with Fig.~\ref{fig:bz}. As expected, $G''$ is suppressed near the antinodes. Panel (c) shows $\Sigma''$, which is insulating-like at PG; at ARC and HS it is metallic. \last{Notice that $\Sigma''$ at PG is enhanced even compared to {its value at} $(\pi,0)$ (dashed red, cf. supplementary note 1).} In the dual formalism the lattice self-energy is given as, $\Sigma_k=\Sigma^{\text{DMFT}}_\nu+\tilde{\Sigma}_k/(1+g_\nu\tilde{\Sigma}_k)$, where $\tilde{\Sigma}$ is the dual self-energy, $g$ is the Green's function of the AIM corresponding to DMFT. At ARC $\Sigma_k''$ is smaller, in absolute value, compared to DMFT. We show that this is the result of nonlocal spin fluctuations.

Fig.~\ref{fig:pg} (d) shows the vertex that couples spin fluctuations with momentum $\Qv$ to fermions at ARC and PG. Note that the imaginary part $\glat''$ is of similar magnitude as the real part $\glat'$. Triangles show the local vertex $\gimp({\nu,\omega=0})$ of the AIM. Its imaginary part is sizable but nonlocal corrections further enhance it. Fig.~\ref{fig:gamma_q} shows $\glat$ for $\nu=\pi T$ as a function of $\qv$ for various dopings. The real part is overall reduced by vertex corrections ($\glat'<1$) and it is suppressed in particular near $\Qv$. This is a precursor to the decoupling of Goldstone excitations from fermions in the antiferromagnet, known as Adler principle~\cite{Adler65,Schrieffer95,Chubukov97,Igoshev07,Huang05,Huang06}; in the extreme case $\xi\rightarrow\infty$ it requires that the vertex vanishes at the ordering vector~\cite{Schrieffer95}. For larger dopings the suppression moves to incommensurate momenta. At PG, the Adler principle does not apply for small doping because of the gap.

\begin{figure}
\begin{center}
\begin{tikzpicture}
\node (image1) at (0,0){\includegraphics[width=0.48\textwidth]{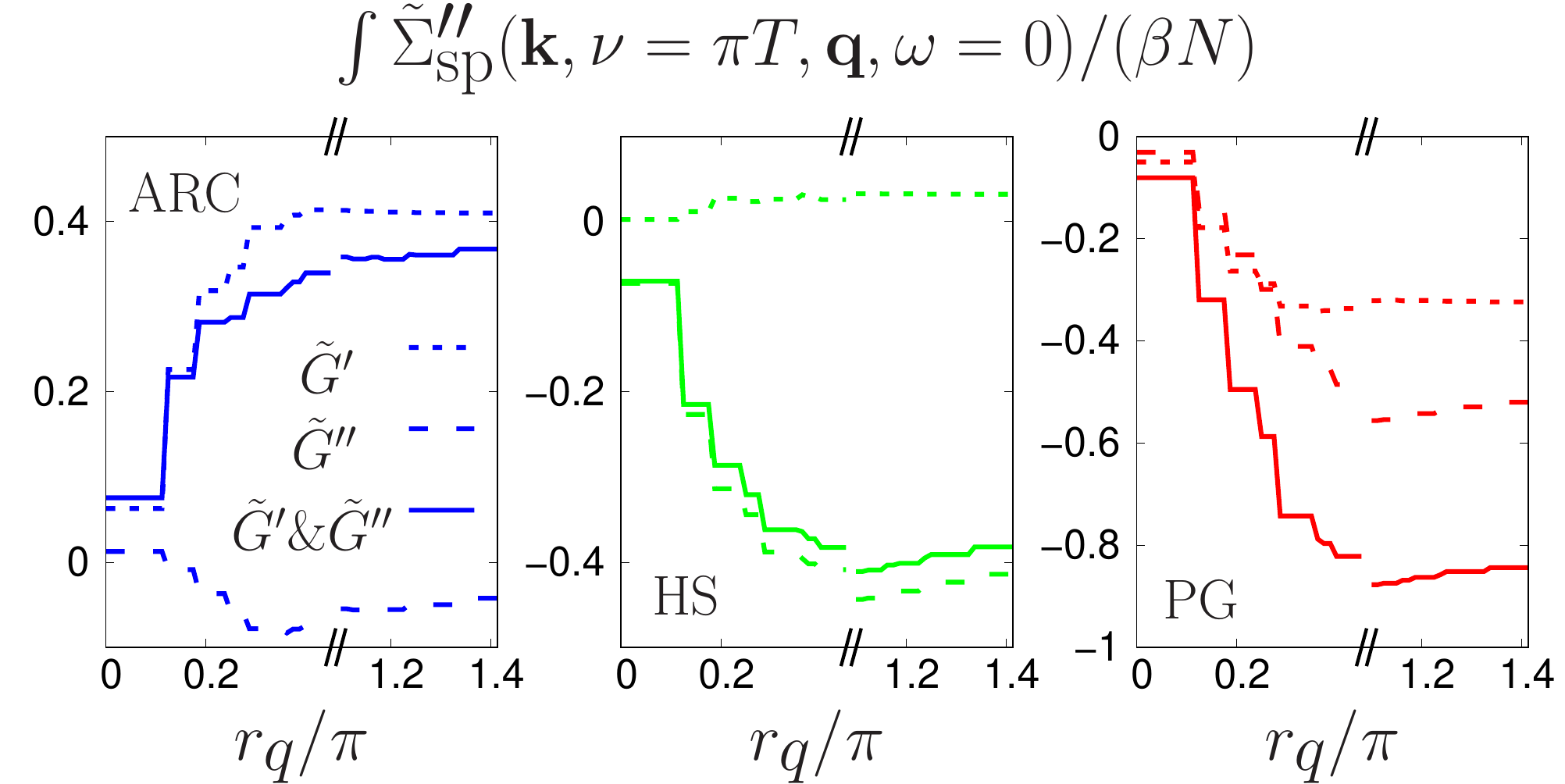}};
\end{tikzpicture}
\end{center}
\vspace{-.5cm}
    \caption{Integrated fluctuation diagnostic (normalized by particle number $N$ and inverse temperature $\beta$) for $\delta=0.01$ as a function of the integration radius $r_q$ (cf. circles in Fig.~\ref{fig:bz}, see text). Dotted and dashed lines show the separate contributions of real and imaginary part of the dual Green's function $\tilde{G}$, respectively.}
\label{fig:diagnostic}
\end{figure}

Figs.~\ref{fig:pg} and~\ref{fig:gamma_q} show that $\glat''$ is large and, hence, the scattering mechanism sketched in Fig.~\ref{fig:bz} needs to be taken into consideration. To reveal its quantitative effect we analyze the contribution $\tilde{\Sigma}_\text{sp}(k,q)$ of nonlocal spin fluctuations to the dual self-energy, it has a form similar to Eq.~\eqref{eq:spinfermion} [cf. supplementary note 1]. First, we integrate $\tilde{\Sigma}''_\text{sp}(\kv,\nu \!=\!\pi T,\qv,\omega\!=\!0)$ with respect to $\qv$ over a circle with radius $r_q$, centered at $\Qv$. This corresponds to circles as in Fig.~\ref{fig:bz}, beginning with $\Qv$ and ending with the entire Brillouin zone~\cite{Krien20-3}. The result is shown as full lines in Fig.~\ref{fig:diagnostic}. A patch of momenta $\qv\approx\Qv$ contributes to the integral, whose final result is negative for PG and HS, but positive for ARC.

This dichotomy can be traced back to $\glat''$. To show this, we split the fluctuation diagnostic into contributions from the real and imaginary part, $\tilde{G}'$ and $\tilde{G}''$, of the dual Green's function. We remind that for $\glat''=0$ the real part $\tilde{G}'$ contributes nothing to the integral for $\tilde{\Sigma}''_\text{sp}$ [cf. Eq.~\eqref{eq:spinfermion}]. Dashed lines in Fig.~\ref{fig:diagnostic} show the contribution of $\tilde{G}''$, which is negative, and absolutely smaller at ARC than at HS and PG. This corresponds to the conventional mechanism which opens a gap near hot spots for $\xi\rightarrow\infty$~\cite{Vilk97,Vilk97-2,Wu18}. Dotted lines in Fig.~\ref{fig:diagnostic} show the contribution of $\tilde{G}'$, which is positive at ARC, negative at PG, and vanishingly small at HS, corresponding to the mechanism sketched in Fig.~\ref{fig:bz}. The pseudogap opens at PG as the  {\sl combined} effect of both mechanisms. Their contributions are comparable at PG, but it is $\tilde{G}'$ ($\Gamma''$) which differentiates the PG from the HS (opens the gap at PG first). With only $\tilde{G}''$ ($\Gamma'$) PG and HS would have similar lifetimes. As already seen in the semi-analytical model, due to $\Gamma''$ non-local spin fluctuations even protect (cool) the ARC ($\tilde{\Sigma}''>0$). We have thus shown that nonlocal spin fluctuations at strong coupling enhance (weaken) correlation effects outside (inside) the AZB.

\begin{figure}
\begin{center}
\begin{tikzpicture}
\node (image1) at (0,0){\includegraphics[width=0.48\textwidth]{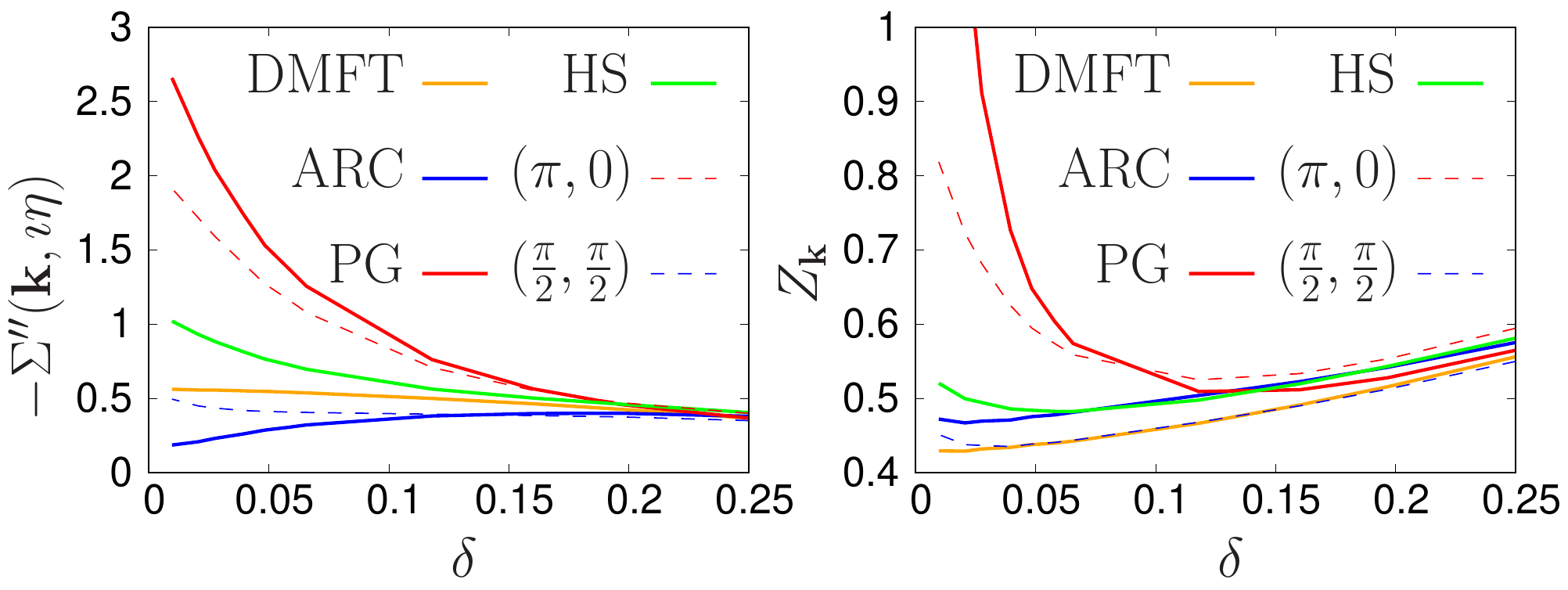}};
\end{tikzpicture}
\end{center}
\vspace{-0.6cm}
    \caption{Scattering rate $-\Sigma''(\kv,\imath\eta)$ and quasiparticle weight $Z_\kv$ versus hole doping $\delta$ at indicated points in the Brillouin zone. \last{Both quantities are obtained through polynomial extrapolation to the \last{Fermi level, $\eta=0^+$}.
}}
\label{fig:scatrat}
\end{figure}

Finally, we explicitly differentiate between scattering rate and quasiparticle weight by \last{extrapolating} the Matsubara self-energy with a fourth-order polynomial \last{to the Fermi level}. The left panel of Fig.~\ref{fig:scatrat} shows $-\Sigma''(\kv,\imath\eta)$ as a function of doping. {As expected, for small dopings the scattering rate is very large at PG, a gap opens and $Z_\kv$, defined through the slope of $\Sigma$, loses its meaning as a quasiparticle weight~\cite{Schaefer21}. At ARC the scattering rate is significantly suppressed compared to DMFT, while $Z_\kv$ remains similar.} Hence, the suppression of the self-energy inside the AZB corresponds primarily to a reduction of the scattering rate (enhancement of the lifetime). The protection of the ARC is so effective that down to $\delta=0.01$ we do not observe the opening of a gap inside the AZB.\newline

\noindent {\textbf{Discussion}}\\
We identified a mechanism for spin-fermion scattering that arises from a combination of strong correlations and particle-hole asymmetry. \last{In the considered temperature and doping regime,}
it dampens quasiparticle excitations on those parts of the Fermi surface
that lie outside of the antiferromagnetic zone boundary,
whereas lifetimes on the inside are actually enhanced by spin fluctuations.
This may explain why the Fermi arcs observed in underdoped cuprates are cut off
at the antiferromagentic zone boundary \cite{ArcExp} and exhibit remarkably good Fermi-liquid properties~\cite{Mirzaei13}. \last{This further indicates that strong non-local correlations cannot simultaneously open an insulating gap on the {entire} Fermi surface.}

{This strong-coupling mechanism \last{is also based on antiferromagentic spin fluctuations, but} it opens the pseudogap  \last{already when} the correlation length is \last{still} smaller than the thermal de Broglie wavelength (for $\delta=0.01$ we estimate $\xi\approx1.6$, while $\xi_{\text{th}}\gtrsim2.1$~\cite{Vilk97}). Nevertheless, only classical spin fluctuations ($\omega=0$) are relevant for the self-energy~\cite{classicalregime}.}

The presented explanation of the strong-coupling \last{spin-fluctuation} mechanism which controls the pseudogap allows us to resolve the contradiction between conventional spin-fluctuation theory and experiments/numerics. As a future perspective, {it is tempting to {also} clarify {its} connection to unconventional superconductivity}. \newline

\noindent {\textbf{Acknowledgments}}\\
We thank H. Aoki, A. A. Katanin, A. Kauch, M. Kitatani, W. Metzner, and A.-M. S. Tremblay for valuable comments and discussions. F.K., P.W., and K.H. acknowledge financial support from the Austrian Science Fund (FWF) through Projects {P32044 and P30997}. P.C.-G. and A.T. acknowledge financial support from the Austrian Science Fund (FWF) Project {No. I 2794-N35}.\newline

\noindent {\textbf{Author Contributions}}\\
{F.K. extended the method of Ref.~\cite{Krien20} to be applicable in the pseudogap regime (cf. supplementary note 1) and performed the calculations. P.W. and P.C.-G. converged DMFT calculations. F.K. and P.W. conceived of the model self-energy and P.W. carried out subsequent calculations.} {F.K., P.W., P.C.-G., A.T., and K.H. analyzed the results and jointly prepared the manuscript.}\newline

\noindent {\textbf{Competing Interests}}\\
{The authors declare no competing interests.}\newline

\noindent {\textbf{Methods}}\\
{The calculations were performed using the boson exchange parquet solver for dual fermions presented in Ref.~\cite{Krien20}. The underlying DMFT problem was solved using continuous-time quantum Monte Carlo solvers with improved estimators~\cite{ALPS2,Hafermann12,Wallerberger19}.}\newline

\noindent {\textbf{Data Availability}}\\
{All data generated during this study are available from the corresponding author on reasonable request (see also supplementary notes 1 and 2).}\newline

\noindent {\textbf{Code Availability}}\\
{All codes used to generate or analyze the results of this study are available from the corresponding author on reasonable request.}
\bibliography{main}

\begin{thebibliography}{82}
\bibitem [{umk()}]{umklapp}%
  \BibitemOpen
  \href@noop {} {}\bibinfo {howpublished} {The importance of the AZB was also
  demonstrated at weak-to-intermediate
  coupling~\cite{Halboth00,Laeuchli04,Honerkamp01,Honerkamp02}, where
  paramagnon scattering leads to a deformation of the
  FS~\cite{Robinson19}.}\BibitemShut {Stop}%
\end{thebibliography}

\end{document}